\documentclass[conference]{IEEEtran}
\IEEEoverridecommandlockouts
\usepackage{cite}
\usepackage{amsmath,amssymb,amsfonts}
\usepackage{algorithmic}
\usepackage{graphicx}
\usepackage{textcomp}
\usepackage{xcolor}
\def\BibTeX{{\rm B\kern-.05em{\sc i\kern-.025em b}\kern-.08em
    T\kern-.1667em\lower.7ex\hbox{E}\kern-.125emX}}
    \usepackage{booktabs}
\usepackage{tabularx}
\usepackage{booktabs}
\usepackage[hidelinks]{hyperref}
\begin{document}

\title{Evaluating IP Blacklists Effectiveness}

\author{\IEEEauthorblockN{Luca Deri}
\IEEEauthorblockA{\textit{ntop} \\
Pisa, Italy \\
deri@ntop.org}
\and
\IEEEauthorblockN{Francesco Fusco}
\IEEEauthorblockA{\textit{IBM Research} \\
Zurich, Switzerland \\
ffu@zurich.ibm.com}
}

\maketitle

\begin{abstract}
IP blacklists are widely used to increase network security by preventing communications with peers that have been marked as malicious. There are several commercial offerings as well as several free-of-charge blacklists maintained by volunteers on the web. Despite their wide adoption, the effectiveness of the different IP blacklists in real-world scenarios is still not clear.

In this paper, we conduct a large-scale network monitoring study which provides insightful findings regarding the effectiveness of blacklists. The results collected over several hundred thousand IP hosts belonging to three distinct large production networks highlight that blacklists are often tuned for precision, with the result that many malicious activities, such as scanning, are completely undetected. The proposed instrumentation approach to detect IP scanning and suspicious activities is implemented with home-grown and open-source software. Our tools enable the creation of blacklists without the security risks posed by the deployment of honeypots.
\end{abstract}

\begin{IEEEkeywords}
IP blacklist, network traffic analysis, host reputation, open-source software.
\end{IEEEkeywords}

\section{Introduction and Motivation}
Reputation systems have been extensively used in network security and network management to maintain networks and networked services secure and reliable. The most commonly used reputation system in network management is blacklisting \cite{oro2010benchmarking, wei2010identifying}. A blacklist is an access control mechanism which denies access to selected network resources to peers belonging to a curated list. Blacklists can be compiled for different resources, including IP addresses, domain names, URLs and checksum of files~(e.g., binary files corresponding to malware). Blacklists often represent the first line of defence for many networks as they can reduce internal hosts' risk of establishing communications with peers with a bad reputation. For example, several anti-spam filters rely on IP reputation blacklists \cite{cook2006catching} which store IP addresses and domain names from which it is not recommended to accept emails.

While IP blacklists \cite{dietrich2009empirical} made the blacklisting approach popular, blacklisting techniques are used at a finer-grained granularity than the IP level, i.e., to block individual network communications marked as suspicious using advanced network fingerprinting mechanisms such as JA3\cite{novakdetection,abuse_ssl}, malicious SSL certificates\cite{ghafir2017malicious} or IoC (Indicators of Compromise), which combine information such as IP addresses, virus signatures and URLs hashes of malware files \cite{bayer2022study,threatfox}.

Blacklists are not used only against malware or spam \cite{zahra2017iot, stevanovic2017method, freudiger2015controlled,oro2010benchmarking, coskun2017wisdom} but also for other purposes such as securing networks \cite{mathur2013detecting, stevanovic2017method} and blocking advertisements \cite{taib2020securing} or improving information delivery and security, including blocking connections from anonymous VPNs or preventing web and security crawlers from scanning a network in search of vulnerabilities that could be potentially used for future attacks~\cite{udger}.

The widespread adoption of IP blacklists has been mostly driven by simplicity and ease of deployment. There are many commercial offerings and several free-of-charge blacklists maintained by volunteers spread across the globe~\cite{emergingthreats,snortip,firehol}. However, when relying on IP blacklists, one has to consider the inherent limitations of the method \cite{ramanathan2020blag}.
First, blacklists are only effective when maintained in a timely manner\cite{xie2007dynamic}. Newly classified malware hosts must be included in the lists, while no longer malicious hosts need to be removed to minimise false positives.
Second, blacklists are not equally effective across the planet. In particular, a blacklist built and maintained for a specific region (e.g., North America) is not guaranteed to be effective when deployed in another region (e.g., Europe). Third, blacklists do not necessarily cover the traffic seen in the network where they are deployed.

Since blacklisting approaches have inherent weaknesses, assessing their effectiveness in real-world scenarios is of extreme importance. In recent years, several rigorous studies have been performed to evaluate and compare malware blacklists~\cite{kuhrer2014paint, essay80567, felegyhazi2010potential}, and domain blacklists when applied to specific application layer services such as email (for spam and phishing) and web traffic~\cite{ thao2018empirical, kuhrer2012empirical, sheng2009empirical}. Surprisingly, despite the availability of a large free-of-charge collection and commercial offerings, there are limited studies on the effectiveness of IP blacklists and many important questions remain unanswered. For example, it is unclear if blacklists are equally effective across distinct networks, which false positive rate has to be expected and which amount of malicious traffic goes undetected.

This paper fills this gap and offers an in-depth study of publicly available IP malware blacklists used in large production networks that are different in size, in nature and span multiple geographical locations over the globe. Our goal is to assess the effectiveness of the blacklists by evaluating them against malicious activities, which can be detected with a high degree of confidence using host instrumentation and aggressive heuristics based on passive network monitoring data. In a nutshell, we wanted to evaluate the effectiveness of the blacklist on the \textit{easy cases}, i.e., for hosts that can be detected as malicious with certainty, even using simple mechanisms. Surprisingly, even compared with our conservative approaches, blacklists can only capture a small fraction of scanning activities, and the recall does not significantly improve when blacklists maintained by distinct parties are combined.

In this paper, we introduce the following contributions. First, we perform a large-scale study evaluating IP blacklists on real-world production networks of more than \textit{hundred thousand} IP hosts belonging to multiple production networks. Second, we describe an effective instrumentation approach to detect IP scanning and suspicious activities toward network servers. Third, we showcase that blacklists are optimized for precision, leaving much of the malicious traffic undetected and a false sense of security. Finally, all the software presented in this paper is home-grown and open-source enabling researchers and network operators to repeat our experiments in their networks.

\section{Data Collection Architecture}
\label{ssec:num0}

To evaluate different IP blacklists, we have selected three networks based in three distinct European locations both using IPv4 and IPv6 addresses:
\begin{itemize}
\item An Italian service provider with about 5'000 hosted servers. This site is connected with two 10 Gbit links to the Internet. Network traffic is mirrored at the edge router, converted into flows by a software probe performing DPI (Deep Packet Inspection) and then collected.
\item A university located in northern Europe with over 100'000 assigned public IP addresses. In this site, network traffic is exported by three border gateways in NetFlow format \cite{claise2004rfc} and collected at a central location.
\item A leading hosting provider located in the Netherlands hosting corporate servers. Each server is monitored using log files (including authentication, web administration, email and TCP/UDP ports monitoring) instead of using raw traffic or flows as in the above two scenarios. These logs contain several invalid login attempts, brute force service attacks, port scans, and data exchange attempts on closed ports. For this reason to the best of our knowledge, no false positives are present, and even if there is a low probability of  misconfigurations or human errors they should be considered very minor and will not change this evaluation. Analysis results are presented in Section~\ref{ssec:num1}. 
\end{itemize}

We have done our best to perform our measurements on heterogeneous networks both IPv4 and IPv6 based, located in different European countries. We are aware that a worldwide measurement system would have been desirable, and we are working at that as reported later in the paper. While our monitoring infrastructure has been active for a longer period of time, in this paper, we report results for three weeks of traffic between February 27th and March 19th, 2023. As the monitoring systems are permanent, data is continuously collected and blacklists are updated daily and matched against the traffic. Over the past five months of traffic monitoring, the results were consistent with what is presented in the paper. It is worth remarking that during the Christmas break, the number of attacks detected decreases significantly compared to non-holiday weeks. For instance, on a mail server, we monitor in this experiment, during the period Dec 27th - Jan 3rd this host received an average of 30k scans/day with a low of 22k on Jan 1st, with respect to 80k scans/day observed in March. A possible interpretation is that behind automated scans there is some sort of human activity that was reduced during holidays, or that perhaps hackers automated the scripts to be less aggressive during holidays.

The main objective of our effort was not to create yet another blacklist to be positioned against the popular ones. Instead, the main goal was to collect the IPs that corresponded to attackers with high confidence in order to evaluate the existing blacklists and understand their strengths and limitations. Our choice to minimize the number of false positives was necessary to estimate:
\begin{itemize}
  \item the number of attackers' IPs present in our network that \textit{are not} present in public IP blacklists.
  \item the false positive rate, i.e., how many IP addresses present in blacklists visited the monitored networks without doing an attack, and thus blocking them would have led to a disservice. This is an important metric for deciding if blacklists can be effectively used as a first line of defence, or if they are unreliable and thus unsuitable for this task.
\end{itemize}

Achieving the goal required analysing the traffic and storing the flow data persistently for the entire period to enable interactive exploration and manual inspection. The matching between the traffic blacklisted and detected was performed daily using the blacklists downloaded on the same day.

\subsection{Network Flow Analysis}
\label{subsect:scanner}
We collect network flows with ntopng~\cite{ntopng}, a popular open-source network traffic monitoring tool we have developed, and store them in ClickHouse~\cite{imasheva2020practice}, a high-performance columnar database \cite{mattis2015columnar}. In addition to providing flow collection capabilities, ntopng implements over hundred distinct rules, which allow the calculation of a \textit{Cyberscore}~\cite{deri2022using} for every network activity. Cyberscore is a scalar indicator that can be computed by combining multiple rules~(called \textit{checks}). Network administrators can configure those checks to identify attackers and the corresponding victims. Since we were interested in minimizing the number of false positives in our setup, we only enabled a handful of simple checks that indicate attacks with very high confidence. The downside of this configuration choice is that some attacks targeting specific hosts might not be considered, but as we will explain later in the section, this is not a limitation. For this reason, ntopng was configured to report alerts produced by remote host scanners, presumably during the pre-attack reconnaissance phase as defined in the Mitre att\&ck \cite{strom2018mitre}, using a simple algorithm: 

\begin{itemize}
\item Similar to \cite{dainotti2012analysis}, we monitor RX traffic of unused IPs in the monitored networks, for which we do not expect to observe incoming traffic. Note that in most networks, not all IP addresses are in use; this happens in particular when monitoring research networks that historically have allocated multiple large (/16 IPv4 or larger) network ranges, and partially used them.
\item Remote peers contacting local receive-only hosts using TCP are reported as scanners if and only if they contact at least 128 receive-only hosts in one day. While TCP is bi-directional by design, UDP traffic is ignored in this computation for two reasons. First, not all UDP protocols are always bidirectional (e.g. Syslog and RTP are not). Second, unidirectional flows using bidirectional UDP protocols such as DNS, NTP, and SNMP are not necessarily an indication of a problem. For instance, SNMP GET is a confirmed PDU, whereas SNMP traps are not. Not considering UDP can reduce the number of scanners we detect, but it protects is from possible false positives for the reasons described above.
\item In order to avoid web crawlers being classified as scanners, traffic on destination TCP ports 80 and 443, are not considered. 
\end{itemize}

We are aware that a more sophisticated mechanism such as a honeypot could be used as attackers targeting websites (for instance those requesting malicious URLs) or scanners limiting their activities to fewer hosts will not be reported (i.e. false negatives). While honeypots could report richer information (for instance it can be used to grab malware software that attackers will drop on the honeypot host) than our passive approach, deploying a honeypot on an operational network is not always possible as many service providers prohibit the installation of software of this nature. Additionally, this conservative approach is not a problem for the blacklist evaluation because with this setup we have very limited, if any. false positives. In all cases, traffic flows are stored in the database and thus retrieved for later analysis if necessary.

\subsection{Server Monitoring}

In addition to identifying scanning activities with the algorithm described in Section~\ref{subsect:scanner}, we monitor log files of core services for the network servers belonging to the networks being monitored. An agent installed on each server monitors both TCP and UDP ports for probing attempts on closed ports, and log files of core services such as SSH, IMAP/SMTP and website management pages. By parsing those logs, the agent can identify intrusion attempts and the IP address from which the attempt has originated. A remote IP is detected as an intruder if logs report multiple unsuccessful access attempts originated at this IP. In addition, similar to what happens with unused hosts in flow monitoring, connections to multiple closed server TCP ports can detect scan attempts, as well as connection attempts to closed ports. The combination of access logs and closed ports monitoring provides evidence of malicious activities, with no false positives, to be computed and used during evaluation together with the scanning detection approach of Section~\ref{subsect:scanner}.


\subsection{IP Blacklist Evaluation}

IP blacklists are available in two formats: as a text file that can be used to implement ACLs (Access Control Lists) on network devices or as cloud services accessible with web REST APIs. As we are interested in evaluating IP blacklists to implement a first level of defence, we need access to them in text format. For this reason, this work focuses on file-based blacklists and uses cloud-based security services to manually verify the results. In particular, we have considered a few popular IP blacklists listed in Table~\ref{table:ip_blacklists}, each with the average number of IP entries. Although all analysed blacklists are designed to identify security threats, some of them have distinctive features. For instance, Feodo tracks only certain types of malware, and Stratosphere \cite{garcia2015modelling} does not track SMTP attacks. These details are relevant for interpreting some of the results presented later in this paper. All blacklists were downloaded for all days of the experiment and evaluated against the observed traffic of the same day. In the rest of this paper, we mainly report only the results for the week of March 13th, with comparable results for all the previous weeks of the experiment. We are unaware of how the lists are created, if they are built in a single geographic region or if they have sources of data spread globally, or what the algorithm is for adding and removing entries. The exception is the Stratosphere Blocklist Generation Project~\cite{stratosphereblocklist}, whose algorithm for the Prioritize New (PN) blacklist is documented in \cite{aip_tool}. Table~\ref{table:ip_blacklists} reports the analysed lists and some details such as the average number of entries and change rate (number of lines that change on average in two consecutive days). As reported, the blacklists' length is not homogeneous as it ranges from a few entries of \textit{dshield} to over 14k of Stratosphere's blacklists. All lists report only IPv4 /32 addresses except for \textit{dshield} and \textit{Emerging Threats} which instead also report subnets. While the work reported in this paper applies and has been evaluated on both IPv4 and IPv6 networks, the evaluation discussed below will be limited to IPv4 as we have not been able to find high-quality and actively maintained IPv6 blacklists.

\begin{table}[ht!]
\caption{Analysed IP Blacklists}
\begin{center}
\begin{tabular}{|l|c|c|c|}
\hline
\textbf{File-based IP blacklists} & \textbf{Entries} & \textbf{IPs} & \textbf{Update Rate}  \\
 &  &  & \textbf{(Daily)}  \\
\hline
\textbf{Snort IP BlockList} \cite{snortip}  & 812  & 812 & 3\%  \\
\textbf{EmergingThreats~(ET)} \cite{emergingthreats} & 1'608 & 16.4 M &  2\% \\
\textbf{Feodo Tracker} \cite{feodo}  & 184 & 184 & 36\% \\
\textbf{dshield}  \cite{dshield} & 29 & 7'936 & 31\%  \\
\textbf{Stratosphere~(PN)} \cite{stratosphere} & 14'518 & 14'518 & 9\% \\
\textbf{AlienVault~(AV)} \cite{alienvault} & 689 & 609 &  1\% \\

\hline
\end{tabular}
\label{table:ip_blacklists}
\end{center}
\end{table}

As shown in Table~\ref{table:ip_blacklists} Stratosphere IPs and EmergingThreats blacklists are the ones containing the highest numbers of IP addresses. ET is the blacklist containing millions of IP addresses due to the use of IPv4 large prefixes such as /16. All blacklists contain only IPv4 addresses.


\begin{table} [ht!]
\begin{center}
    \caption{Analysed Cloud-based IP Blacklists}
    \begin{tabularx}{0.8\columnwidth}{|X|}
        \hline
        \textbf{Cloud-based IP blacklists} \\
        \hline
        \textbf{VirusTotal} \cite{virustotal}  \\
        \textbf{AbuseIP DB} \cite{abuseipdb} \\
        \textbf{Greynoise} \cite{greynoise} \\
        \hline
    \end{tabularx}
    \label{table:ip_blacklists_cloud}
    \end{center}
\end{table}

While nothing can be said about cloud-based blacklists shown in Table~\ref{table:ip_blacklists_cloud}, for file-based ones, Table~\ref{table:ip_blacklists} reports the number of entries and the percentage of entries that change daily. Some blacklists update entries very seldom, whereas others are more dynamic. Most lists are updated daily with more changes during the middle of the week rather than on week-ends, others (AlienVault) are updated less frequently. Before comparing blacklists against the collected network data, blacklists were cross-compared to test the intersection of IP addresses among them. The comparison was implemented using a Python script that uses a Patricia tree \cite{pytricia} module to interpret IP addresses.

\begin{table}[ht!]
\caption{Blacklists IP Entries Intersection (Contains/Contained)}\label{tab:ugly}
\begin{center}
\begin{tabular}{|l|c|c|c|c|c|c|}
\hline
 List Name & \textbf{PN} &  \textbf{AV}  &  \textbf{Snort}  &  \textbf{ET}  &  \textbf{Feodo} &  \textbf{dshield} \\
\hline
\textbf{PN} & - & 3 & 13 & 0 & 0 & 0\\
\textbf{AV} & 3 & - & 0 & 0 & 0 & 0\\
\textbf{Snort} & 13 & 0 & - & 1 & 0 & 0\\
\textbf{ET} & 978 & 1 & 3 & - & 170 & 10\\
\textbf{Feodo} & 0 & 0 & 170 & 0 & - & 0\\
\textbf{dshield} & 1029 & 0 & 0 & 11 & 0 & - \\
\hline
\end{tabular}
\label{table:blacklists_intersection}
\end{center}
\end{table}

Table~\ref{table:blacklists_intersection} contains the intersection of the lists on March 13th, with similar results (i.e. minor differences in value less than 10\%) for the other days of the experiment. The intersection of the lists (i.e. an IP address that is included in the intersection of two lists if it is contained in both of them) is limited. The results are biased for lists that contain subnets as it is unlikely that all subnet IPs are reported in other lists. This is because we assume that a subnet entry is fully included in another list when all the subnet IPs are included, and it explains for instance why in the above table PN (that uses /32 prefixes) has 978 entries contained in ET (that uses larger subnets), but not the opposite as no ET network completely fits inside the PN network list.


\subsection{Comparing Flow Data with File-based IP Blacklist}

\begin{figure}[htbp]
\centerline{\includegraphics[width=0.8\columnwidth]{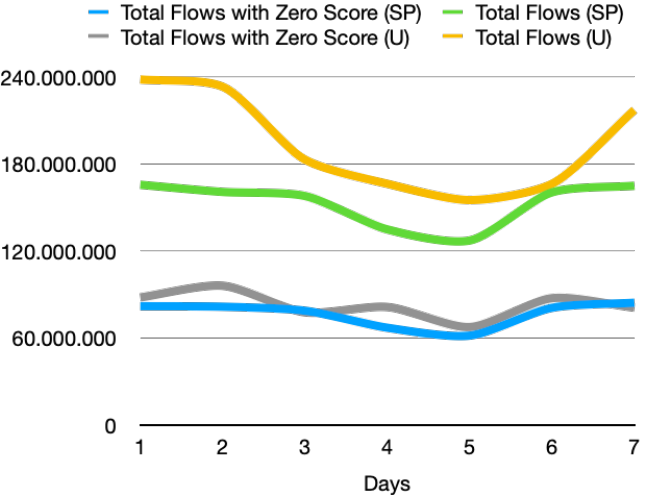}}
\caption{Flows Distribution (March 1st-7th)}
\label{fig1}
\end{figure}

Figure \ref{fig1} shows the flow distribution of our experiments for the first 7 days of March 2023 at the service provider (SP) and university (U) network.

Table~\ref{table:dataset_size} reports the dataset size, including the number of IPs with zero cyberscore as defined in \ref{subsect:scanner}.

\begin{table}[ht!]
\caption{Daily Dataset Size (7 Days Average)}
\begin{center}
\begin{tabular}{|l|c|c|c|}
\hline
 & \textbf{Service}  & \textbf{University} \\
 & \textbf{Provider}  & \textbf{Network} \\
\hline
\textbf{Total Flows} & 153M  & 194M \\
\textbf{Total Flows with Zero Cyberscore} & 76M & 82M\\
\textbf{Active Local IPs} & 49K & 79K \\
\textbf{Unique Remote Client IPs} & 1.7M & 71K \\
\textbf{Remote Scanner IPs} & 1.8K & 3.1K\\
\textbf{Remote IPs with Zero Cyberscore} & 1.3M & 65K\\
\hline
\end{tabular}
\label{table:dataset_size}
\end{center}
\end{table}

Only 6\% scanner IPs of Table~\ref{table:dataset_size} visited both networks on the same day. These numbers change significantly when monitoring the scanners over two weeks.

\begin{figure}[htbp]
\centerline{\includegraphics[trim = {0 0cm 0 1.0cm}, clip,width=0.8\columnwidth]{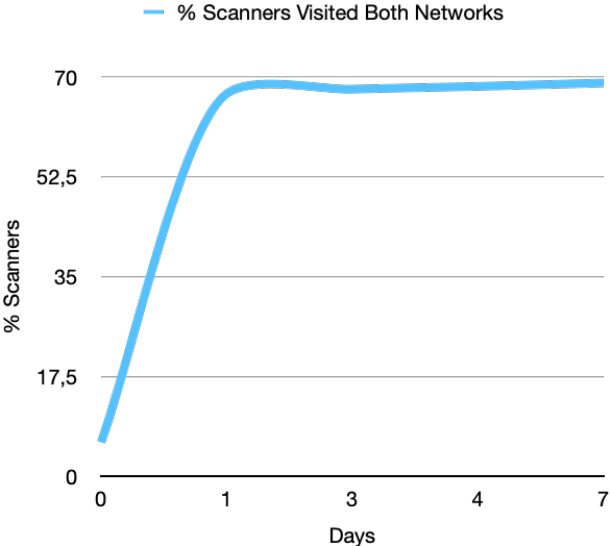}}
\caption{Scanners Propagation (7 Days)}
\label{fig2}
\end{figure}

Figure \ref{fig2} shows that after one day 67\% of the university scanners also visited the service provider network. After 13 days this percentage slowly increased to 70\%. The interpretation of this data is that most scanners move across networks in less than a day, so promptly updating lists is a key property that a blacklist should have to be effective. In addition, having a large set of monitoring nodes across the Internet helps create a comprehensive blacklist that includes attackers' IPs that soon will visit the network to protect. Neighbourhood Watch is an old concept used by humans, that demonstrated to be effective as an early warning system \cite{nw_university} for Internet monitoring, and thus that should also be considered in blacklists. When some hosts are under scan or attack, likely the attackers will also target other hosts belonging to the same network: if attacker IPs are immediately blacklisted, this could be a very effective early-warning technique to protect the whole network.

Another interesting finding is the average number of unique client IP addresses that visited both networks in a single day. This is probably justified by the fact that even if the service provider network is much smaller, due to the nature of this business, most hosts are servers, whereas, at the university, it is probably the opposite. This said, about 6.5k unique client IP addresses visited both networks on the same day. Meaning that even if these two networks are different, there is some limited overlap. When looking at these overlapping IPs more in detail, if the last IP address byte is ignored (essentially merging them in /24 networks), we accounted for 20k networks with at least 8 IP addresses, 18k networks with 8...63 IP addresses, and 1.5k networks with at least 64 IP addresses. Randomly selecting IPs from networks with at least 8 IPs and searching them on an IP reputation database \cite{thalos}, most of these networks have a poor reputation or are used by companies that have as core business periodically scan the Internet (e.g., \textit{internet-measurement.com} and \textit{shodan.io}).

In Table~\ref{table:blacklists_vs_flows}, we report the average number of IP addresses corresponding to scanners that are present in the blacklists. For convenience, we report the percentage of IP addresses present in each blacklist. What appears evident from the results is that the gap between blacklists is substantial. Providers such as AlienVault, Feodo and Snort are ineffective in marking scanner IPs. Stratosphere Prioritize New is the best blacklist to identify scanner IPs, and augmenting it with all the other blacklists improves the detection only marginally (by 1\%).

\begin{table}[ht!]
\caption{Scanner IPs Match Rate with File-based IP Blacklists (7 Days Average)}\label{tab:ugly}
\begin{center}
\begin{tabular}{|l|c|c|}
\hline
\textbf{Blacklist} & \textbf{Service Provider} & \textbf{University} \\
\hline
\textbf{Stratosphere (PN)} & 50.5\% & 14.6\% \\
\textbf{EmergingThreats (ET)} & 13.6\% & 4.8\%\\
\textbf{dshield} & 11.3\% & 4.5\% \\
\textbf{AlienVault} & 0\% & 0.1\%  \\
\textbf{Snort} & 0.1\% & 0\%  \\
\textbf{Feodo} & 0\% & 0\%  \\
\textbf{PN+ET+dshield} & 50.9\% & 14.7\% \\
\hline
\end{tabular}
\label{table:blacklists_vs_flows}
\end{center}
\end{table}

We have tried to evaluate the effectiveness of blacklists over time to understand how blacklist updates affect results. For this reason, we have matched the traffic daily (i.e. at midnight of day X we download the blacklist that will match the traffic of the day) against using the blacklist of day X for matching the traffic of days X+1, X+2...X+5 and results of 5 days comparison are reported in  Table~\ref{table:pn_service_provice_decay}. As expected, using a daily blacklist produces slightly better results than reusing the same blacklist for matching scanner IPs of the following days. However, the slow decay rate is probably surprising, i.e. blacklist of day X is also effective for a couple of days more with no match rate degradation.

\begin{table}[ht!]
\caption{Stratosphere AP Matches (Mar 13-17th) using Mar 13th Blacklist (Service Provider)}\label{tab:ugly}
\begin{center}
\begin{tabular}{|l|c|c|c|c|c|}
\hline
\textbf{ Match} &  \textbf{Mar 13} & \textbf{Mar 14} &  \textbf{Mar 15}  &  \textbf{Mar 16}  &  \textbf{Mar 17} \\
\textbf{ Rate} & & & & &   \\
\hline
\textbf{ Daily (\%)}       & 49\% & 53\% & 51\% & 51\% & 51\% \\
\textbf{ Delta (\%)} & 0\% & 0\% & -0.3\% & -0.5\% & -1.4\%  \\
\hline
\end{tabular}
\label{table:pn_service_provice_decay}
\end{center}
\end{table}

In order to evaluate if blacklists contain false positives, we have extracted the list of IPs whose flows had a maximum cyberscore of 100 (minor issues). We have matched such IPs against all the file-based blacklists and found no overlap, i.e. no flow client IP was listed in blacklists. This experiment reported similar results for the university network. This result is very encouraging as it indicates that blacklists seem to have no false positives with flows having no or minor issues. Said that a more in-depth analysis of flows with a higher score should be performed, the above result encourages the utilization of blacklists for blocking traffic.

In another test, we evaluated the IP blacklist coverage, namely, out of all malicious communications detected by ntopng (including but not limited to suspicious DGA domain contacted, possible exploit, and TLS certificate mismatch) performed by remote peers, what is the (daily) percentage of peers listed in one of the analysed IP blacklists. ntopng can trigger over 100 different types of cybersecurity checks covering both plain-text and encrypted traffic analysis \cite{deri2021using}. To avoid false positives and make this experiment reproducible we have enabled only the default checks that trigger alerts based on nDPI security risks (nDPI is an open-source deep packet inspection library on top of which ntopng is built): nDPI inspects the initial flow content and produces a flow risk assessment based the flow packet payload. We have extracted from the alert database the list of remote peers that performed malicious activities from monitored traffic as described in Section~\ref{ssec:num0}. This has allowed us to identify both network and port scanners listed above, as well as other types of threats reported by nDPI that we have used for validating the blacklists. This assessment is based on the assumption that nDPI has a very low number of false positives if any \cite{radityatama2017study, bujlow2013comparison}, and to the best of our knowledge this is the case being it is used in many firewall and IPS tools that use it to block traffic and thus that cannot tolerate false positives. In nDPI, the security analysis is based on flow payload inspection (e.g. malformed packet, expired or self-signed TLS certificate, invalid HTTP request) and flow state (e.g. TCP flow with no answer) without enabling checks based on heuristics that might be affected by false positives. We have considered only those hosts that ntopng considers malicious with high confidence (i.e. host cyberscore bigger than 5'000).

\begin{table}[ht!]
\caption{ntopng Alerted IPs (Match \%) Listed in File-based IP Blacklists (Service Provider)}\label{tab:ugly}
\begin{center}
\begin{tabular}{|l|c|c|c|c|c|c|}
\hline
\textbf{Day} & \textbf{PN} &  \textbf{AV}  &  \textbf{Snort}  &  \textbf{ET}  &  \textbf{Feodo} &  \textbf{dshield} \\
\hline
\textbf{Mar 13}        & 37\% & 0\% & 0\% & 6\% & 0\% & 7\% \\
\textbf{Mar 14}        & 33\% & 0\% & 0\% & 8\% & 0\% & 13\% \\
\textbf{Mar 15}        & 34\% & 0\% & 0\% & 7\% & 0\% & 3\% \\
\textbf{Mar 16}        & 38\% & 0\% & 0\% & 9\% & 0\% & 7\% \\
\textbf{Mar 17}        & 44\% & 0\% & 0\% & 7\% & 0\% & 13\%  \\
\textbf{Mar 18}        & 24\% & 0\% & 0\% & 2\% & 0\% & 3\% \\
\textbf{Mar 19}        & 31\% & 0\% & 0\% & 0\% & 0\% & 0\% \\
\textbf{Average 13-19} & 34\% & 0\% & 0\% & 6\% & 0\% & 7\% \\
\hline
\end{tabular}
\label{table:all_client_ip_matchrate_listed}
\end{center}
\end{table}

Table~\ref{table:all_client_ip_matchrate_listed} shows that in the best case, less than 50\% of the hosts were present in blacklists, which confirms the results reported in Table~\ref{table:blacklists_vs_flows} that consider only scanners. For those not having a match in the blacklist, we have taken a random subset of 500 hosts (as explained later this is the daily limitation for two of the considered IP blacklists). Over 60\% of these IPs are listed in cloud IP blacklists. The outcome is that file-based blacklists can detect only a subset of scanners and attackers, and therefore while they are still useful, file-based IP blacklists are not sufficient to successfully protect a network.


\subsection{Comparing Flow Data with Cloud-based IP Blacklist}

As anticipated earlier in this paper, we have compared cloud-based blacklists with file-based blacklists. Cloud-based services periodically aggregate information from various sources (VirusTotal). Others label an IP as malicious if there is a minimum consensus (AbuseIP), and others (Graynoise) instead use other proprietary and undisclosed techniques. As these services allow a limited number of searches per day (50 hosts/week for GreyNoyse, 500/day for VirusTotal and 1000/day for AbuseIP with free plans), we have selected two 500 IPs lists (one observed in the service provider and the other in university network) scanners addresses from a single day of the observation period scanners' IP list and checked the match rate using both file (of the same day) and cloud-based blacklists. For GreyNoyse the experiment was limited to 50 hosts due to the above limitations.

\begin{table}[ht!]
\caption{Cloud Based Blacklists Match Rate: 500 IP Scanners (Service Provider / University)}\label{tab:ugly}
\begin{center}
\begin{tabular}{|l|c|c|c|c|}
\hline
\textbf{Blacklist} & \multicolumn{2}{c|}{\textbf{Service Provider}} & \multicolumn{2}{c|}{\textbf{University}} \\
\cline{2-5}
& List. & Cons. & List.& Cons.\\
\hline
\textbf{VirusTotal} & 75\% & 80\% & 62\% & 63\%  \\
\textbf{Abuse IP DB} & 98\% & 25\% & 68\% & 18\%\\
\textbf{Greynoise (50 IPs)} & 34\% & 10\%  & 34\% & 10\%  \\
\textbf{PN (File-based)} & 75\% & 12\%  & 75\% & 12\%  \\
\textbf{ET (File-based)} & 36\% & 4\%  & 36\% & 4\%  \\

\hline
\end{tabular}
\label{table:cloud_based_500_random_ips}
\end{center}
\end{table}

The table above shows the results of this experiment, including the match rate of file-based blacklists on the same IPs. Results are divided into two columns: the first one reports a match if the IP is listed, and the second only if there is also a consensus (5 or more matches for VirusTotal, 100\% accuracy for Abuse IP, and 'malicious' for GreyNoise). Repeating the experiment on a different day of the observation period produces similar results. It is odd to report that except VirusTotal, all other blacklists perform poorly when matching scanner IPs coming from the university network. According to the above results, with the limitations of the number of queries per day, cloud-based blacklists seem to outperform file-based ones significantly. In conclusion, the nature of cloud-based services prevents them from being used to dynamically configure ACLs in security devices. Therefore, these cloud services are helpful for security analysts but cannot be practically used as the first line of defence as file-based blacklists.


\subsection{Comparing Server Monitoring Data with IP Blacklist}
\label{ssec:num1}

This section evaluates intrusion attempts on two hosts, one used as an email and the other as a web server. The same service provider hosts both servers, however on two very different networks located in the Netherlands. Only two administrators from specific IP addresses have access to the server, and to avoid interference with collected data, their accesses have not been considered in the analysis using the servers as honeypots. All attempts are recorded by looking at multiple (3+) failures in authentication and application (e.g. email and web server) log files and connection attempts on closed TCP ports. Contrary to the previous experiment where ntopng was used, here we have used {ipt\_geofence} (\url{https://github.com/ntop/ipt_geofence}) which is deployed on the monitored hosts. This is an open-source tool we have developed and is able to detect scanning attempts, contacts to closed TCP/UDP ports and service attacks looking at system log files (e.g. invalid login reported in the auth.log file). For this reason, we assume that all the above activities represent real scans or non-authorised failed connection attempts not affected by false positives; this is as we take into account only multiple failures on servers used only by the author. Table~\ref{table:cloud_based_500_random_ips} reports the results of a single day of analysis in the paper observation period with similar results recorded on the other days of the week of March 13th.

\begin{table}[ht!]
\caption{Server Monitoring Data Analysis}\label{tab:ugly}
\begin{center}
\begin{tabular}{|l|c|c|}
\hline
\textbf{Attackers} & \textbf{Web} & \textbf{Mail} \\
& \textbf{Server} & \textbf{Server} \\
\hline
\textbf{Total number of Attacks} & 75 & 450 \\
\textbf{In Blacklist} & 73\% & 42\% \\
\textbf{Also visited Service Provider} & 14\% & 11\% \\
\textbf{Also visited University} & 5\% & 4\% \\

\hline
\end{tabular}
\label{table:cloud_based_500_random_ips}
\end{center}
\end{table}

Out of this experiment, these are the main findings:
\begin{itemize}
  \item Blacklists are more effective for the web server host rather than the mail server, even if the total number of attacks on the mail server is higher. We are not sure why this happens for all blacklists except Stratosphere which is a list built without any email malware feeds. Probably the fact that the two servers are hosted on two totally different networks, and the different intent of blacklists is relevant here.
  \item Comparing attackers with the list of IPs that visited the same-day service provider and the university network, we observe that only a small minority visited such networks. In this case, as observed before, attackers are visiting the service provider more often than the university. This is a bit odd as the university network has many more hosts than the service provider. Instead, looking at the data over 7 days, we have observed that 70\% of the scanners also visited the other network which implies that probably it is just a matter of time. We need to study our data over a longer period of time with more monitored hosts in order better study this fact.  
  \item Another fact worth remarking is that while both hosts are dual stack, 99\% of the attackers use IPv4 addresses.
  
\end{itemize}


\section{Conclusions}

In this paper, we describe our experience in evaluating free-of-charge IP blacklists over two large production networks and Internet corporate servers. We propose an effective method which can be used to identify attackers with high confidence, with the goal of estimating the coverage of IP blacklist. The method combines passive monitoring with log analysis which is performed directly on production servers using local agents. Our large-scale study shows that IP blacklists provide limited protection against attackers. In particular, top free-to-use blacklists identify about 50\% of the scanner attacks. Additionally, we find out that even if different blacklists are not strongly overlapping, i.e. their intersection is low, by combining all the blacklists the coverage can be only marginally improved. The low recall of the malicious IPs can be explained by the small false positive rate, indicating that blacklists are most likely optimized for precision rather than recall.
Cloud-based blacklists can help to confirm malicious activities, but the limitation in the API queries prevents these types of blacklists from being used as the first line of defence. 

Our methodology in evaluating IP blacklists suggests that detection algorithms tuned for precision (i.e. they include only hosts that performed only major malicious activities in order to minimise false positives, as the blacklists we have evaluated have limited coverage of attacks detected in our experiments) when used in large geographically distributed and always-on monitoring infrastructures might represent a valuable strategy to improve current blacklists.


\section{Future Work}

This paper monitored traffic in  different European locations. A future goal is to extend this work to non-European sites and different network types including residential and mobile networks.

The experiments described in this paper have shown that blacklists are affected by two main limitations: they only detect a subset of network scanners, and their detection rate is not constant across the service provider and the university networks. We believe that deploying more sensors across the Internet, better if in heterogeneous networks, could address the above issues. This is the driving force for creating a feed for publicly available IP malware blacklists using the scanning detection mechanisms developed in this paper (and eventually more sophisticated ones). We are already developing a prototype that we would like to start deploying in the coming months.

Finally, the number of IPs in the blacklists that never incurred malicious activities against the monitored infrastructure was not addressed in this study but represents a topic of interest to better understand the quality of the blacklists in general, and thus a future work item.

\section*{Acknowledgment}

The authors would like to thank Joaquin Bogado, and Sebastian Garcia for endless discussions, experiments evaluation and suggestions throughout this research work. 

\bibliographystyle{plain}
\bibliography{references}

\end{document}